\documentclass[reprint,superscriptaddress]{revtex4-1}
\usepackage{amsmath,graphicx}
\begin{document}

\title{Femtometer-resolved simultaneous measurement of multiple laser wavelengths in a speckle wavemeter}

\author{Graham D. Bruce} \email{gdb2@st-andrews.ac.uk}
\affiliation{SUPA School of Physics and Astronomy, University of St Andrews, North Haugh, St Andrews KY16 9SS, UK}

\author{Laura O'Donnell}
\affiliation{SUPA School of Physics and Astronomy, University of St Andrews, North Haugh, St Andrews KY16 9SS, UK}
\author{Mingzhou Chen}
\affiliation{SUPA School of Physics and Astronomy, University of St Andrews, North Haugh, St Andrews KY16 9SS, UK}
\author{Morgan Facchin}
\affiliation{SUPA School of Physics and Astronomy, University of St Andrews, North Haugh, St Andrews KY16 9SS, UK}
\author{Kishan Dholakia}
\affiliation{SUPA School of Physics and Astronomy, University of St Andrews, North Haugh, St Andrews KY16 9SS, UK}
\affiliation{Department of Physics, College of Science, Yonsei University, Seoul 03722, South Korea}

\begin{abstract}
Many areas of optical science require an accurate measurement of optical spectra. Devices based on laser speckle promise compact wavelength measurement, with attometer-level sensitivity demonstrated for single wavelength laser fields. The measurement of multimode spectra using this approach would be attractive, yet this is currently limited to picometer resolution. Here, we present a method to improve the resolution and precision of speckle-based multi-wavelength measurements. We measure multiple wavelengths simultaneously, in a device comprising a single 1\,m-long step-index multimode fiber and a fast camera. Independent wavelengths separated by as little as 1\,fm are retrieved with 0.2\,fm precision using Principal Component Analysis. The method offers a viable way to measure sparse spectra containing multiple individual lines and is likely to find application in the tracking of multiple lasers in fields such as portable quantum technologies and optical telecommunications. 
\end{abstract}

\maketitle

The speckle produced when coherent light is scattered by a rough surface can provide a surprising method with which one can track the properties of the incoming light. The precise speckle pattern produced by this multiple-interference is uniquely determined by the beam parameters, and can therefore be used as a fingerprint for linewidth \cite{Freude86}, polarization \cite{Kohlgraf10}, beam position \cite{Alexeev17} or transverse mode characteristics \cite{Mazilu12}. Broadband spectrometers have been constructed which extract the spectrum of light from the speckle, by using either the transmission matrix method (\cite{Redding13,Redding14,Chakrabarti15,Cao17}) or deep learning \cite{Kurum19}, achieving a spectral resolution limited by speckle correlation. Typically, this speckle correlation limit is on the picometer-scale. For monochromatic light, speckle wavemeters utilizing Principal Component Analysis (PCA) \cite{Mazilu14,Metzger17,Bruce19}, Poincar\'e descriptors \cite{ODonnell19} and convolutional neural networks \cite{Gupta19} have greatly surpassed this limit, measuring an isolated wavelength with a resolution down to the attometer-scale. It remains an open challenge to simultaneously measure multiple wavelengths or spectra at such high resolution using speckle. A successful method promises applicability in laser stabilization for portable cold atoms experiments, wavelength-division multiplexed telecommunications and chemical sensing.

In this letter, we demonstrate that the high resolution achieved by using PCA to analyze speckle can be extended beyond a single laser-line, to measure sparse spectra composed of multiple laser wavelengths. We establish that wavelength measurements of lasers separated by 1\,fm, five orders of magnitude less than the speckle correlation limit, can be performed simultaneously and with an accuracy of 0.2\,fm. Simultaneous measurement of up to ten laser lines is demonstrated.

\begin{figure}[htbp]
\centering
\includegraphics[width=1\linewidth]{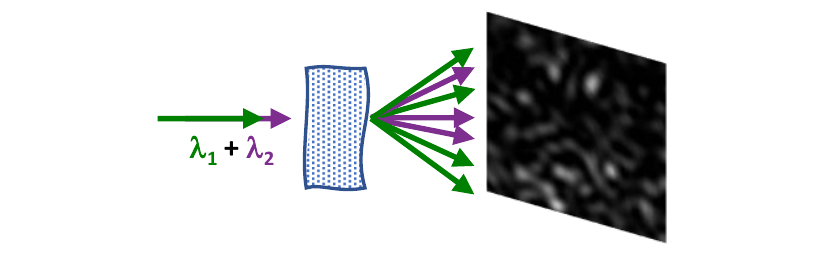}
\caption{Principle of multi-wavelength measurement in a speckle wavemeter. Laser beams are overlapped and illuminate a single scattering medium, generating a speckle pattern. The speckle pattern is uniquely determined by the precise values of each wavelength, so can be used as a marker to recover the wavelengths. \label{fig0}}
\end{figure}

The principle of measurement is outlined in Fig.~\ref{fig0}. A single scattering element is illuminated by a beam composed of multiple wavelengths; each wavelength is scattered to produce a unique speckle pattern. Provided the wavelengths of the Components are sufficiently separated, the resultant speckle patterns are a simple intensity sum of the speckles produced by each wavelength in isolation. A calibration dataset is acquired to train PCA to recognize how the speckle changes with wavelength.

\begin{figure}[t]
\centering
\includegraphics[width=1\linewidth]{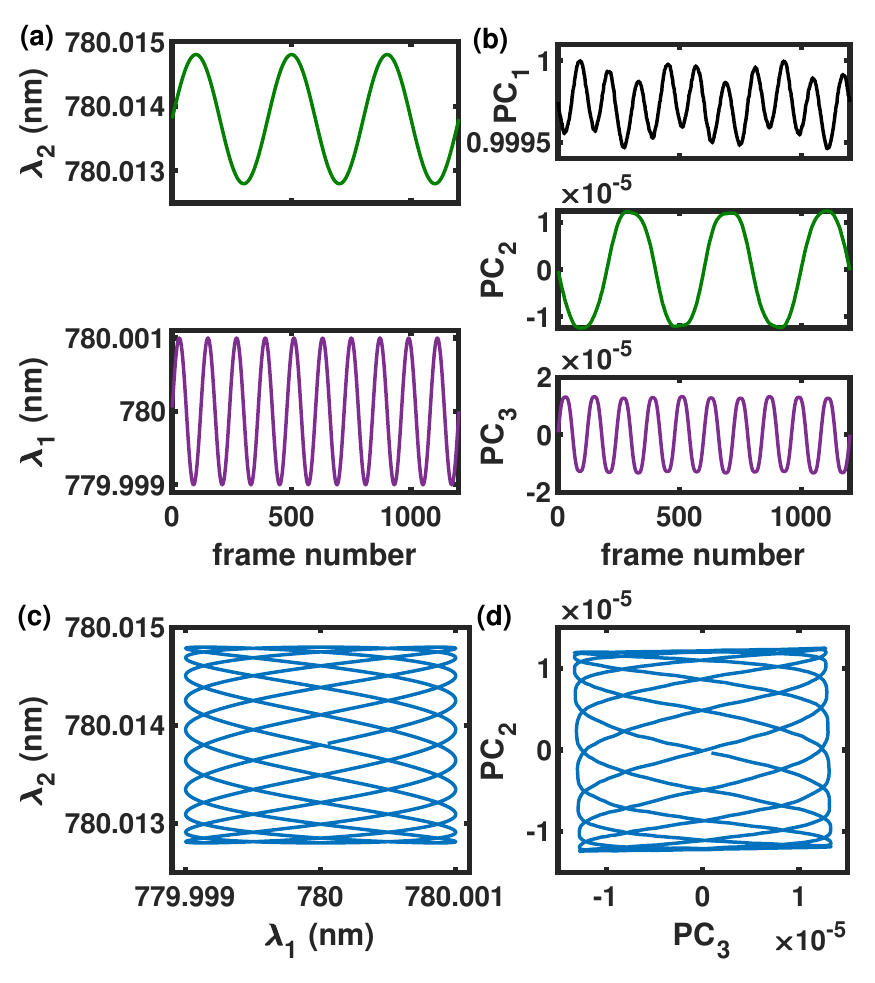}
\caption{Principal Component Analysis of simulated speckle patterns produced by wavelength variations of two overlapped lasers. (a) Sinusoidal modulations applied to the wavelengths of two lasers $\lambda_{1}$ and $\lambda_{2}$. (b) Principal Components 1 - 3 of the image set. The first Principal Component (PC$_{1}$) captures a mixed signal of both wavelength modulations, while PC$_{2}$ and PC$_{3}$ show responses dominated by $\lambda_{2}$ and $\lambda_{1}$ respectively. Parametric plots of (c) $\lambda_{1}$ vs $\lambda_{2}$ and (d) PC$_{3}$ vs PC$_{2}$ show that the combined modulations are faithfully recorded in PC-space. A small rotation angle between (c) and (d) highlights mixing of the two wavelength Components across the two PCs. \label{fig1}}
\end{figure}

\begin{figure}[!t]
\centering
\includegraphics[width=1\linewidth]{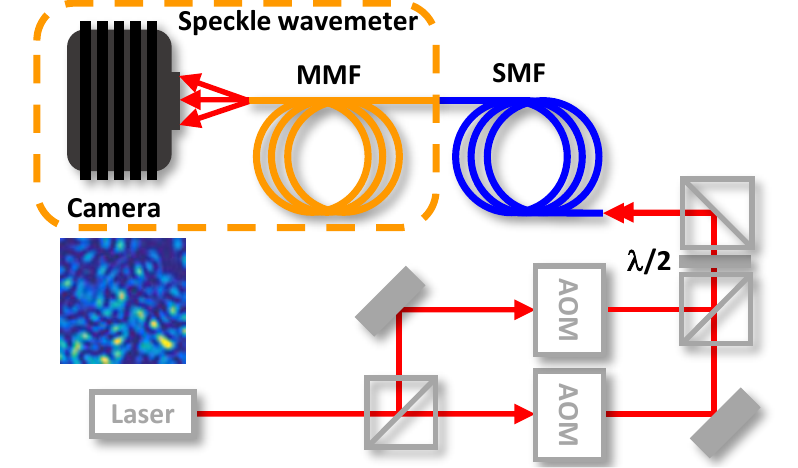}
\caption{Experimental setup. The output of a stabilized diode laser is split into two beams, each of which undergoes separate wavelength modulation using an acousto-optic modulator (AOM). The beams are recombined, co-polarized and delivered to the speckle wavemeter via single-mode optical fiber (SMF). The speckle wavemeter is a 1\,m-long multi-mode fiber (MMF) and a CMOS camera. (Inset) A typical multi-wavelength speckle pattern recorded in the speckle wavemeter. 
\label{fig2}}
\end{figure}

To demonstrate the training method, we simulate  (using paraxial wave theory, see \cite{Metzger17} for details) the propagation of two co-polarized, co-incident and co-propagating Gaussian laser beams of equal power and identical spatial distribution. The light propagates through five equally-spaced planes (separated by one Rayleigh length) at which the phase is randomized. The refractive index difference to air is small ($\Delta n=0.001$) to ensure most scattering is in the forward direction. After further free-space propagation of two Rayleigh lengths after the final randomization, the resulting speckled intensity is sampled on a $256\times256$ pixel grid with a bit-depth of 8, to approximate the acquisition by a camera. A series of 1200 speckle patterns are accumulated, where wavelengths $\lambda_{1}$ and $\lambda_{2}$ of the two lasers are centered around 780.000\,nm and 780.014\,nm. They are both sinusoidally modulated with a 1\,pm-amplitude but with different periods of oscillation (such that they undergo three and ten oscillations in the measurement period, respectively), as shown in Fig.~\ref{fig1}(a). At each time interval, the multi-wavelength speckle pattern is obtained by summing the intensities of the speckle distribution of each wavelength in isolation, i.e. neglecting interference between the two beams. PCA is then performed on the time-series of multi-wavelength speckle patterns. The Principal Components (PCs) are the projections of the data onto the eigenbasis of the covariance matrix of the training set, i.e. by design they measure the maximal variations in the dataset. The largest three PCs (PC$_{1}$, PC$_{2}$ and PC$_{3}$, shown in Fig.~\ref{fig1}(b)), contain 96\% of the variations in the data. The non-commensurate modulation rates for the two beams ease the identification of the contribution from each wavelength. The first Principal Component, PC$_{1}$, shows modulation of the speckle pattern at both of the applied modulation rates. This is associated with intensity fluctuations due to speckles moving in and out of the field of view of the camera. However, the separate modulations are dispersed across the next two Principal Components (PC$_{2}$ and PC$_{3}$ in Fig.~\ref{fig2}(b)), in analogy with the wavelength-dependent dispersion produced in a grating-based spectrometer. Retrieval of these two PCs in isolation is sufficient to characterize the independent wavelengths: the parametric relationship between $\lambda_1$ and $\lambda_2$ is illustrated in Fig.~\ref{fig1}(c), and the same parametric relationship is shown to exist between PC$_{2}$ and PC$_{3}$ in Fig.~\ref{fig1}(d). A small rotation angle between the two parametric plots signifies cross-talk between the two measurement channels, i.e. PC$_{2}$ is strongly dependent on $\lambda_2$ and weakly dependent on $\lambda_1$ while PC$_{3}$ is strongly dependent on $\lambda_1$ and weakly dependent on $\lambda_2$. We find that this cross-talk can be minimized by using wavelength modulations of equal amplitude, but regardless it does not effect the accuracy of the PCA, as the two wavelengths are always uniquely identified by the measurement of these two PCs. The link between PCs and wavelength is established by a linear fitting of this training set. A speckle pattern produced by an unknown combination of wavelengths within the training range can subsequently be projected into this PC-space to retrieve the wavelengths.

We experimentally verify the method using the apparatus shown in Fig.~\ref{fig2} to generate tunable, multi-wavelength spectra. Light from an external cavity diode laser (Toptica DL-100, LD-0785-P220), stabilized to the $^{87}$Rb D$_2$ line ($F=2 \rightarrow F’=2\times3$ crossover) with saturated absorption spectroscopy and current modulation, is separated into two beams using a polarizing beam splitter. The wavelength of each beam is shifted by independent acousto-optic modulators (AOMs) (Crystal Technologies 3110-120) in cat-eye double-pass configuration, with a modulation range for each beam of 20\,fm. The two beams are recombined and co-polarized using further polarizing beam splitters and a half-waveplate. The light is coupled into an angle-cleaved single-mode fiber (SMF) (ThorLabs P5-780PM-FC-10) to ensure each beam has the same spatial profile, and delivered to a multi-mode fiber (MMF) speckle wavemeter. Laser speckle is generated by multiple scattering and modal interference in the 1\,m-long step-index MMF, which has 105\,$\mu$m core diameter and NA = 0.22 (ThorLabs FG105LCA). After exiting the MMF, the light propagates for 5\,cm and is captured by a fast CMOS camera (Mikrotron EoSens 4CXP). Images of $240\times240$ pixels were recorded at 2,000\,fps with an exposure time of 10\,$\mu$s and a power of 150\,$\mu$W per beam. The multi-wavelength speckle image at each time interval is independently normalized by the total intensity. The speckle correlation limit of this system is $\sim320$\,pm, which is determined as the HWHM of the Pearson correlation coefficient of the speckle patterns at different wavelengths.

\begin{figure}[!t]
\centering
\includegraphics[width=1\linewidth]{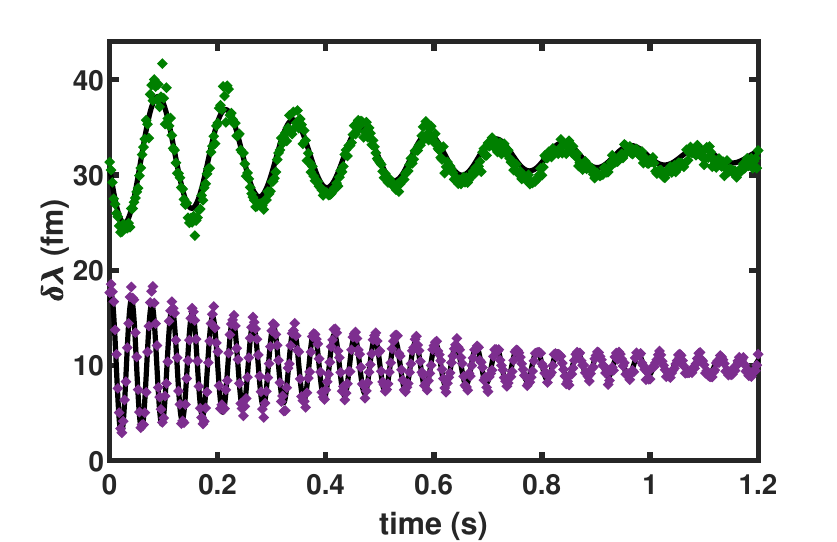}
\caption{Simultaneous measurement of two wavelengths, shown relative to $\lambda_{0} = 780.2437$\,nm. The black line denotes the control modulation applied to the AOM, and the points denote the retrieval of wavelength from the speckle wavemeter. \label{fig3}}
\end{figure}

Fig.~\ref{fig3} shows the measurement of two wavelengths with an average separation of 22\,fm, which is four orders of magnitude below the speckle correlation limit and for which the speckle patterns acquired at each wavelength have a structural similarity index $>0.97$. Training was performed by acquiring the speckle patterns over a 1\,s interval for a 8.5\,fm-amplitude sinusoidal wavelength modulation to each beam, with incommensurate periods of 125\,ms and 37.5\,ms. After the training phase, an exponential decay of the amplitude of the wavelength modulation is introduced. The standard deviation between the set wavelength and that measured by the speckle wavemeter is 0.37\,fm for the slowly modulated beam and 0.29\,fm for the fast modulated beam. The accuracy of the measurement of each wavelength is limited by high-frequency modulations of the laser wavelength introduced by the lock-in electronics for wavelength stabilization, and are in agreement with those reported in \cite{Bruce19} for measurements of a single wavelength.

When the wavelength separation is large, PCA accurately recovers the wavelength. However, if the wavelengths converge, PCA is incapable of correctly analyzing the speckle pattern, giving large values of the PCs which do not correspond to the expected wavelengths. The erratic values for close approach are due to interference between the beams causing the speckle pattern on the camera to flicker. When the beat-note frequency of this interference-induced flicker is fast compared to the exposure time of the camera, PCA gives reliable wavelength estimation. Using the camera settings above, we measured the wavelengths of two beams to an accuracy (standard deviation of measured and set wavelength over 1\,s) of 0.21\,fm and 0.19\,fm when the wavelength separation was 1.0\,fm. In principle, decreasing the measurement rate and using longer exposure times should allow for an improvement in spectral resolution, while the issue may be avoided in the measurement of separate laser sources. 

In addition to wavelength separation, we also investigated the role of other potential issues with our approach. For modest power ratio between the beams, the two wavelengths are always uniquely determined by PC$_{2}$ and PC$_{3}$. However, when this power ratio is large, e.g. $500\mu$W and $50\mu$W, the less intense beam is instead dispersed into PC$_4$, therefore further PCs must be considered to track multiple wavelengths in this regime. When the lasers have different linewidth, the differing Rayleigh distributions of the resultant speckle patterns will aid the discrimination of the contributions of each laser to the speckle.

\begin{figure}[!t]
\centering
\includegraphics[width=1\linewidth]{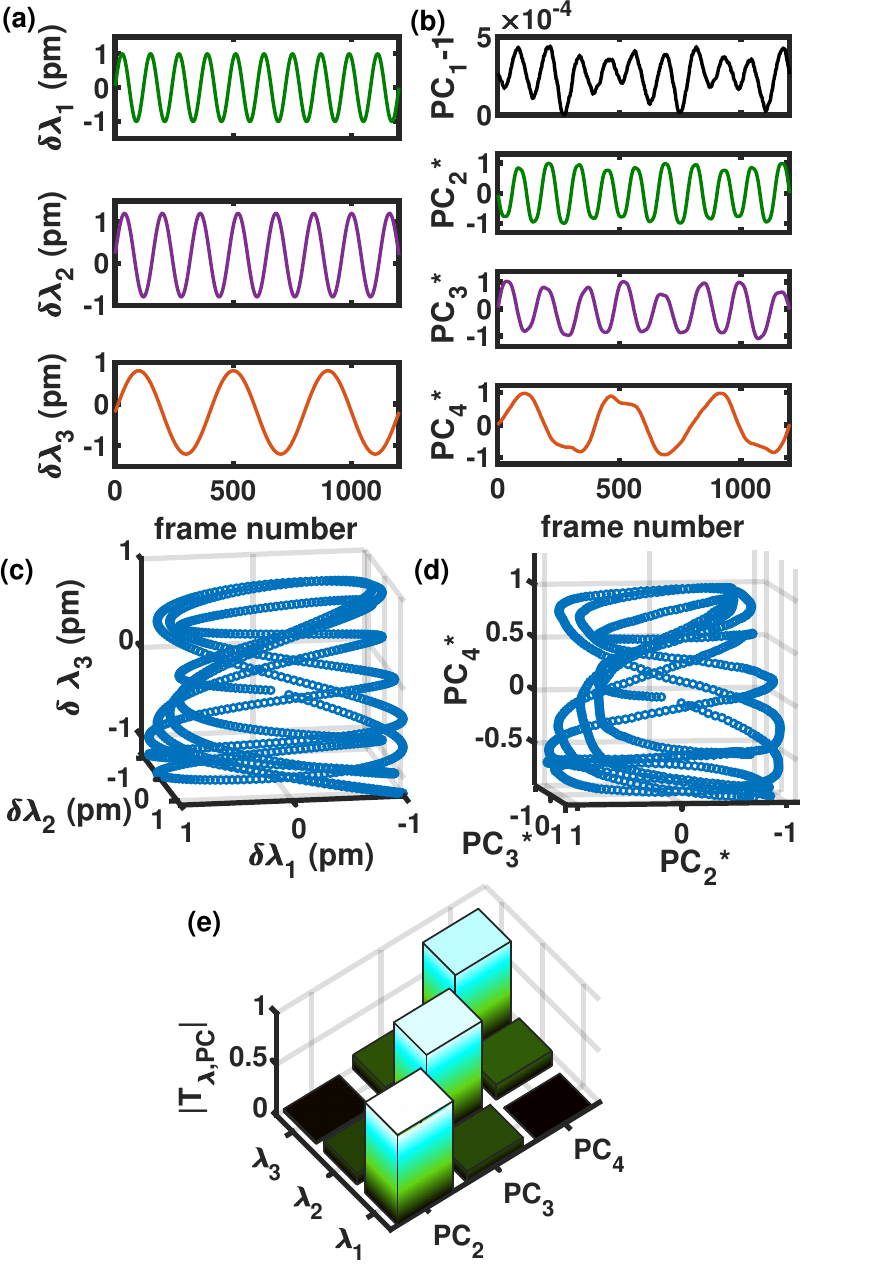}
\caption{Tracking three wavelengths simultaneously via Principal Component Analysis. (a) Control wavelength modulations to the three lasers. $\delta\lambda_{1-3}$ are measured relative to 780\,nm, 780.986\,nm and 779.014\,nm respectively. (b) Principal Components 1 - 4 of the resultant speckle patterns. PC$_{i}$* denotes PC$_{i} \times 10^{5}$. Parametric plots (c) of wavelengths and (d) of PCs, showing that the PC-space representation is related to the wavelength-space by a three-dimensional rotation. (e) The transformation matrix $T_{\lambda,\text{PC}}$ gives the relationship between each wavelength and each PC.  \label{fig4}}
\end{figure}

Simultaneous measurements of more than two wavelengths are also possible. Fig.~\ref{fig4}(a) shows simulated wavelength modulation of three separate beams, with mean separations of 14\,pm, which are combined as before. As in the two-wavelength case, the wavelength modulations are dispersed across PC-space (97\% of the variation is described by the first four PCs). PC$_{1}$ shows a mixture of all three modulations, while PC$_{2}$ to PC$_{4}$ are respectively dominated by $\lambda_1$ to $\lambda_3$ (Fig.~\ref{fig4}(b)). Mixing of the spectral channels is again observed: the 3-dimensional parametric plot of wavelength (Fig.~\ref{fig4}(c)) is related to the parametric plot of PC$_{2}$, PC$_{3}$ and PC$_4$ (Fig.~\ref{fig4}(d)) by a 3-dimensional rotation. The mixing of spectral channels can also be seen in the transformation matrix $T_{\lambda,\text{PC}}$ which defines the linear transformation between PCs and wavelength, i.e.
\begin{equation}
    \begin{pmatrix}
    \text{PC}_{2} \\
    \text{PC}_{3} \\
    \text{PC}_{4}
    \end{pmatrix} = 
    \begin{pmatrix}
    T_{\lambda_{1},\text{PC}_{2}} & T_{\lambda_{1},\text{PC}_{3}} & T_{\lambda_{1},\text{PC}_{4}} \\
    T_{\lambda_{2},\text{PC}_{2}} & T_{\lambda_{2},\text{PC}_{3}} & T_{\lambda_{2},\text{PC}_{4}} \\
    T_{\lambda_{3},\text{PC}_{2}} & T_{\lambda_{3},\text{PC}_{3}} & T_{\lambda_{3},\text{PC}_{4}}
    \end{pmatrix}
    \begin{pmatrix}
    \lambda_{1} \\
    \lambda_{2} \\
    \lambda_{3}
    \end{pmatrix}.
\end{equation}
\noindent $T_{\lambda,\text{PC}}$ is established in the training phase by multiplication of the matrix containing the time series of the PCs and the inverse of the matrix containing the time series of the corresponding training wavelengths, and is plotted in Fig.~\ref{fig4}(e).  It shows that the mean dependence of $\text{PC}_{i+1}$ on $\lambda_{i}$ is 83.1\%, with an average contribution of 12.1\% from the nearest neighboring wavelength(s). The wavelengths present in any individual unknown speckle pattern can be measured by multiplying the matrix inverse of $T_{\lambda,\text{PC}}$ with the PCs extracted for that image.

\begin{figure}[!t]
\centering
\includegraphics[width=1\linewidth]{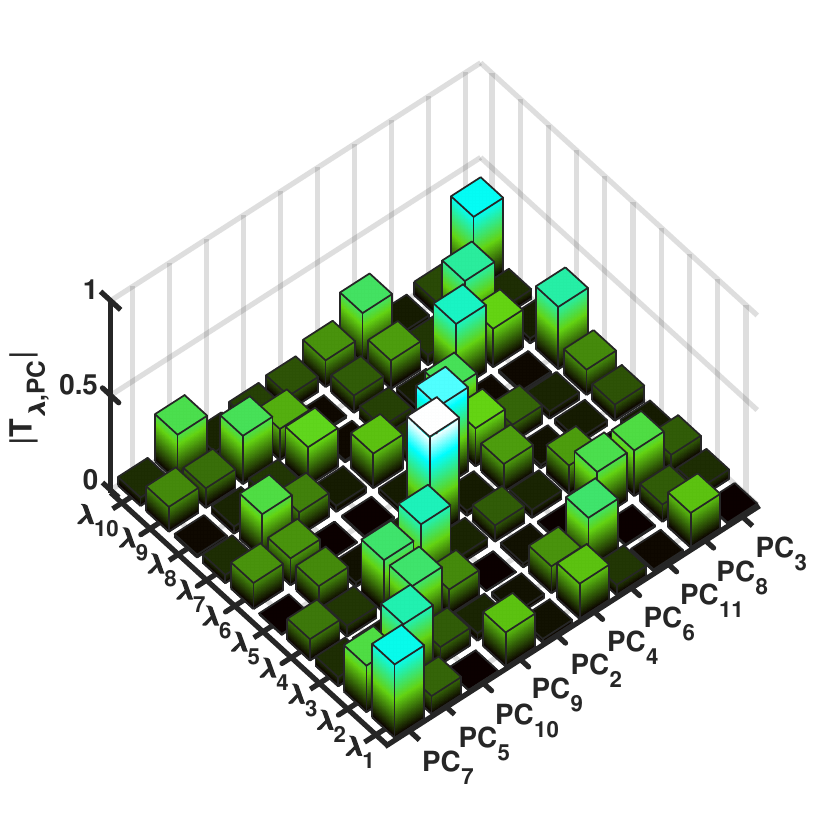}
\caption{Transformation matrix $T_{\lambda,\text{PC}}$ showing the relationship between each wavelength and each PC in the speckle patterns produced by ten overlapped wavelengths.  \label{fig5}}
\end{figure}

The transformation matrix representation is necessary to examine the correlations between higher numbers of beams. As shown in Fig.~\ref{fig5}, a similar transformation matrix can be established for a system of 10 distinct lasers, where the ten wavelengths are dispersed across PC$_{2}$ to PC$_{11}$. In this simulation, the ten wavelengths were evenly separated by 1\,pm, and undergo incommensurate sinusoidal modulations of 200\,fm amplitude over 400 frames. The period of oscillation of $\lambda_{i}$ was set so that it undergoes $2 p_{i}$ oscillations in the training phase, where $p_{i}$ is the $i$th prime integer. The PCA finds a basis in which 74\% of the variance is contained in the first eleven PCs. We note that the variance captured in higher PCs in this case follows a step-like trend in groups of ten PCs: continuously falling by 50\% within the group but discontinuously dropping by 50\% between the last PC of one group and the first PC of the next. Ignoring these higher terms and projecting test data into the 10-dimensional PC space comprising PC$_{2}$ to PC$_{11}$ recovers the wavelength to within 20\,fm. As can be seen in Fig.~\ref{fig5}, there is greater mixing between the spectral channels in this ten-wavelength measurement, with the diagonal elements of $T_{\lambda,\text{PC}}$ having a mean value of 31.3\% and a standard deviation of 9.1\%. 

In this letter, we have demonstrated that the wavelengths of multiple lasers can be measured simultaneously using a speckle wavemeter with Principal Component Analysis. The procedure projects a speckle pattern generated by $n$ wavelengths into an $n$-dimensional Principal Component space. In the experiment, we demonstrated simultaneous recovery of the wavelengths of two lasers, separated by as little as 1\,fm with an accuracy of 0.2\,fm, limited by the stabilization electronics of the laser. The approach is limited in spectral range, requiring that the Principal Components vary monotonically with wavelength. However, for single wavelength measurements, PCA has been shown to be complimentary to the transmission matrix method, which operates over a much larger range but with lower resolution \cite{Metzger17}. We suggest such a tandem approach will also be possible for the measurement of multiple wavelengths. The method is likely to find application in the development of portable quantum technologies, where robust methods to lock multiple lasers for atom cooling are sought. Couturier, et al, have shown that such stabilization can be achieved using a commercial (Fizeau) wavemeter and a multi-mode fiber switch, but report fluctuations of the atomic fluorescence due to the switching \cite{Couturier18}. Stabilization of a single laser using speckle was demonstrated in \cite{Metzger17}, and we suggest that the simultaneity of measurements of multiple wavelengths with speckle may obviate the switching limitation. In future work, the training phase could be extended to include variable powers of the beams, which would allow for the recovery of sparse spectra with variable mode intensities, which may be applicable to areas such as chemical analysis.

We acknowledge funding from Leverhulme Trust (RPG-2017-197)and  EPSRC (EP/R004854/1). 
and thank D Cassettari and P Rodr\'iguez-Sevilla for technical assistance and discussions.

\bibliography{sample}

\end{document}